\documentclass[aps,pra,showpacs,amssymb]{revtex4}
\usepackage{graphicx}
\usepackage{bm}
\begin{document}
%\LARGE
%\preprint{gr-qc/}
%\draft
\title{Specific heat of a particle on the cone}
\author{E. S. Moreira, Jr.}
 \email{moreira@unifei.edu.br}
\author{E. S. Oliveira}
 %\email{}
\affiliation{Instituto de Ci\^encias Exatas,
Universidade Federal de Itajub\'a, 
Av.\ BPS 1303 Pinheirinho, 37500-903 Itajub\'a, MG, Brazil} 

\date{March 21, 2006}
%\twocolumns[
%\hsize\textwidth\columnwidth\hsize\csname@twocolumnfalse\endcsname 

\begin{abstract}

This work investigates how a conical singularity can affect
the specific heat of systems.
A free nonrelativistic particle confined to 
the lateral surface of a cone --- conical box ---
is taken as a toy model.
Its specific heat is determined
as a function of the deficit angle and the temperature.
For a vanishing deficit angle, the specific heat is that of a 
particle in a flat disk where a characteristic temperature separates
quantum and classical behaviors, as usual.
By increasing the deficit angle the characteristic
temperature increases also, and eventually 
another characteristic temperature 
(which does not depend on the deficit angle) arises.
When the cone gets sufficiently sharp, at low and intermediate temperatures
the azimuthal degree of freedom is suppressed.
At low temperatures the specific heat varies discontinuously with the
deficit angle. Connections between certain theorems regarding  common
zeros of the Bessel functions and this discontinuity
are reported.

\end{abstract}

\pacs{03.65.Ge, 02.30.Gp, 05.20.-y, 05.30.-d}

\maketitle

%\begin{multicols}{2}

\section{Introduction}

As techniques in condensed matter physics
to fabricating very thin samples with various 
topologies and geometries develop, 
the study of quantum mechanics on curved surfaces
becomes increasingly relevant.
Among these backgrounds a cone is perhaps the
one with the simplest nontrivial geometry --- it is flat
everywhere except at its tip 
where there is a conical singularity \cite{mar59,sta63,dow67,sok77,des84}.
Due to local flatness 
the conical singularity does not affect local dynamics; but it
does affect global features
such as the solutions of differential equations.
One is dealing, therefore, with a geometric analog
of the  Aharonov-Bohm setup \cite{bez89}.

There has appeared a considerable number of works on
quantum mechanics of a particle in conical spaces, 
since the early investigations in Refs. \cite{dow77,tho88,lan90}. 
Nevertheless, it seems that a study on 
how a conical singularity affects the 
statistical mechanics of a particle  has not yet been pursued.
It should be mentioned that
since physics around a topological 
defect in the bulk of a solid can be seen as physics in an 
effective conical geometry \cite{kro81},
the study of statistical mechanics in conical spaces may help to learn
about thermal properties  when topological defects are present.

This paper addresses quantum statistical mechanics 
of a free nonrelativistic particle on the cone. 
The focus is on the specific heat $C$.
In Sec. II the classical partition function is calculated, showing that 
classical $C$  is not affected by the conical singularity.
In Sec. III, the particle energy spectrum is obtained and an 
analysis on how it depends on the deficit angle is 
presented. In particular, connections with properties of the zeros
of the Bessel functions are discussed.
In the following section, the energy spectrum is used to obtaining 
the quantum specific heat. Quantum $C$ is then studied at low, 
intermediate and high temperatures,
showing that the conical
singularity can affect it substantially. 
In Sec. V the paper is closed with a summary and final remarks.

\section{Classical statistical mechanics}

One begins by defining the background
(a more detailed account on the conical geometry can be found
for example in Refs. \cite{des84,mor98}). A wedge of angle ${\cal D}$
is removed from a disk of radius $a$, 
and the two opposite sides of the missing wedge are
identified, resulting a cone of deficit angle ${\cal D}$.
It is convenient to express ${\cal D}$ in terms of the cone parameter
$0<\alpha\leq 1$,
\begin{equation}
{\cal D}=2\pi(1-\alpha).
\label{cparameter}
\end{equation}
Then a flat disk (i.e., absence of conical singularity)
corresponds to $\alpha=1$,  
and the cone becomes increasingly sharper
as $\alpha$ decreases from unity.
It is worth mentioning that a cone with ${\cal D}=\pi$
corresponds to the configuration space of a system of
two identical particles (here separate by a maximum distance $a$) 
with respect to its centre of mass \cite{lei77}.

A free norelativistic particle of mass $M$ 
in thermodynamical equilibrium with a reservoir at
temperature $T$ is constrained to move on the cone described above. 
The corresponding Hamiltonian is given by
\begin{equation}
\label{chamiltonian} 
{\cal H}=\frac{p_{\rho}^{2}}{2M}
+ \frac{p_{\varphi}^{2}}{2M\rho^{2}},
\end{equation}
where $p_{\rho}$ and  $p_{\varphi}$
are the usual momenta associated with polar coordinates 
$\rho$ and $\varphi$.
%\begin{eqnarray}
%0\leq \rho \leq a &\hspace{1.0cm}& 0 \leq \varphi \leq 2\pi\alpha.
%\label{ranges}
%\end{eqnarray}
Although Eq. (\ref{chamiltonian}) 
carries no trace of the conical singularity
[Eq. (\ref{chamiltonian}) is simply the Hamiltonian of a free particle 
on the Euclidean plane], the latter appears in the  identification
\begin{equation}
(\rho,\varphi+2\pi\alpha)\sim (\rho,\varphi).
\label{identification}
\end{equation}
It can be shown that the conical singularity at $\rho=0$ 
corresponds to an integral curvature 
proportional to ${\cal D}$ \cite{sok77}.

Observing Eq. (\ref{chamiltonian}), 
the standard procedure to compute the (semi) classical 
partition function
by integrating over the phase space \cite{pat72},
\begin{equation}
Z_{(2)}=\frac{1}{h^{2}}\int_{0}^{a}d\rho\
\int_{0}^{2\pi\alpha}d\varphi
\int_{-\infty}^{\infty}dp_{\rho}
\int_{-\infty}^{\infty}dp_{\varphi}\
e^{-{\cal H}/\kappa T},
\label{cpfunction}
\end{equation}
yields
\begin{equation}
Z_{(2)}=\frac{A}{h^{2}}2\pi M\kappa T,
\label{cpfunction2}
\end{equation}
where
\begin{equation}
A:=\alpha \pi a^{2}
\label{area}
\end{equation}
is the area of the cone. 
Once a partition function $Z$ is known, the internal energy is
given by
\begin{equation}
U=\kappa T^{2}\frac{1}{Z}\frac{dZ}{dT}.
\label{ienergy}
\end{equation}
As the dependence
of $Z_{(2)}$ on $\alpha$ happens only through a 
temperature independent factor, 
the internal energy
and, consequently, the specific heat
\begin{equation}
C=\frac{dU}{dT}
\label{sheat}
\end{equation}
are clearly not affected by the conical singularity. 
Indeed, these formulae give
\begin{eqnarray}
U=\kappa T &\hspace{1.0cm}& C=\kappa
\label{ienergy2}
\end{eqnarray}
as one might have guessed by evoking the equipartition theorem
at an early stage [see Eq. (\ref{chamiltonian})].
[In fact, as a straightforward covariant calculation can show,
the results in Eqs. (\ref{cpfunction2}) and (\ref{ienergy2}) 
hold for any two-dimensional curved box regardless its topology and geometry.]

\section{Energy spectrum}
In order to use the tools of quantum statistical mechanics,
one needs to know the energy spectrum of the system ---
a free particle in a conical box.
Considering Eq. (\ref{identification}) 
the stationary states $\Psi$ must satisfy  
$$\Psi(\rho,\varphi+2\pi\alpha)=\Psi(\rho,\varphi).$$
Eq. (\ref{chamiltonian}) leads to a Hamiltonian operator which is
simply $-\hbar^{2}/2M$ times the flat Laplacian written in polar
coordinates. One then faces the task of solving
the eigenvalue problem for the flat Laplacian operator 
on the cone
\cite{tho88,lan90}, with the additional
boundary condition 
$$\Psi(\rho=a,\varphi)=0,$$
since the particle is confined to a circular  
box of radius $a$. (This is essentially the eigenvalue problem 
which appears in the study of vibrations of a circular membrane
\cite{gra66}.) 

Up to a normalization constant, the regular solutions 
(i.e., finite at $\rho=0$) are
\begin{equation}
\Psi=J_{|n|/\alpha}(k_{n,s}\rho)\exp\{in\varphi/\alpha\}\hspace{2.0cm}
k_{n,s}:=\frac{1}{a}\chi_{|n|/\alpha,s}
\label{states}
\end{equation}
where 
$n$ is an integer, $s$ is a positive integer, and
$\chi_{\nu,s}$ denotes the $s$th positive zero of the 
Bessel function
of the first kind $J_{\nu}(x)$ \cite{gra66,wat44,abr65,gra80}. 
Approximated values for the zeros can be obtained by truncating 
the expansion \cite{gra66}
\begin{equation}
\chi_{\nu,s}=
\left(\frac{\nu}{2}+s-\frac{1}{4}\right)\pi
-\frac{4\nu^{2}-1}{8\pi}
\left(\frac{\nu}{2}+s-\frac{1}{4}\right)^{-1}-\ \ldots
%(\nu/2+s-1/4)\pi
\label{avalue}
\end{equation}
at some point which depends on the accuracy required.
Eq. (\ref{avalue}) is specially adequate in dealing with high zeros
(large $s$). 

Then, for a pair of quantum numbers
$(n,s)$,  $\Psi$ is a state with angular momentum $n\hbar/\alpha$ 
($n>0$ and $n<0$ correspond to the particle winding 
counterclockwise and clockwise, respectively) 
and energy
\begin{equation}
E_{n,s}=\frac{\hbar^{2}k_{n,s}^{2}}{2M},
\label{energy}
\end{equation}
both quantities, therefore, discrete. 
As $E_{n,s}=E_{-n,s}$, $E_{n\neq 0,s}$ is at least doubly degenerate
(degeneracy will be addressed in more detail below).

In order to have a picture of the energy spectrum, 
and of the corresponding stationary states, one recalls
that $J_{\nu}(x)$ is an oscillatory function of $x$ and that its zeros
$\chi_{\nu,s}$ are labelled such that 
$\chi_{\nu,1}<\chi_{\nu,2}<\chi_{\nu,3}< \ldots$.
It is also important  to know that $\chi_{\nu,s}$ is a continuous increasing function
of $\nu$ \cite{wat44}.
Then $E_{n,s}\geq E_{0,1}$, i.e.,
\begin{equation}
E_{0,1}=\frac{\hbar^{2}}{2M}\left(\frac{2.4048\ldots}{a}\right)^{2}
\label{zpenergy}
\end{equation}
is the zero point energy
($\chi_{0,1}$ has been replaced by its numerical
value \cite{abr65}), and $E_{n,s}$ increases with $|n|/\alpha$ and $s$. 
Clearly, energy levels with $n=0$ 
are not affected by the conical singularity
(even classically the conical singularity only affects the motion if the
particle has nonvanishing angular momentum \cite{mor98}).

To figure out the way that  $\alpha$ [${\cal D}$, cf. Eq.(\ref{cparameter})]
shapes the energy spectrum,
it helps to know that for $\alpha=1$ 
(particle confined to a flat disk \cite{rob02})
the energy spectrum resembles
that of a particle in a square box of the same area, 
although this analogy should not
be pushed too far (e.g., the degeneracies of the corresponding energy 
levels do not match completely).
When $\nu > 0$,
$\chi_{\nu,s}>\nu$ \cite{gra80}, leading to  
\begin{equation}
E_{1,1}>\frac{\hbar^{2}}{2Ma^{2}\alpha^{2}}.
\label{einequality}
\end{equation}
Since $E_{1,1}$ is the energy of the first excited state with $n\neq 0$,
as the cone gets sharper (i.e., as $\alpha$ decreases from unity) 
all the energy levels with
nonvanishing angular momentum are swept up in the spectrum,
leaving those with vanishing angular momentum unchanged. 
It should be mentioned that for large $\nu$, 
$\chi_{\nu,1}\simeq \nu +1.8557\nu^{1/3}$ \cite{abr65}, resulting
that ``$>$'' in Eq. (\ref{einequality}) can be replaced by 
``$\simeq$" when $\alpha$ is small,
with increasing accuracy as $\alpha\rightarrow 0$.

In order to discuss degeneracy of the energy levels, it is appropriate
to begin with $\alpha=1$. 
As two Bessel functions of the first kind 
with integer orders cannot have common zeros
(this follows from a theorem known as Bourget's hypothesis \cite{wat44}),
it results that
energy levels with $n=0$
are not degenerate, whereas those with $n\neq 0$
are doubly degenerate. 
This situation changes when the conical singularity is present
($\alpha\neq 1$), leading to a new source of degeneracy, since the 
Bessel functions in Eq. (\ref{states}) may have common zeros 
(see Ref. \cite{pet03} and references therein).

Noting that 
$\chi_{0,1}<\chi_{1,1}<\chi_{2,1}<\chi_{0,2}<\ldots$ \cite{abr65},
when $\alpha=1$ Eq. (\ref{energy}) yields
\begin{equation}
E_{0,1}<E_{1,1}<E_{2,1}<E_{0,2}<\ldots .
\label{a1}
\end{equation}
As $\alpha$ decreases,  $E_{1,1}$ and $E_{2,1}$ increase,
whereas $E_{0,1}$ and $E_{0,2}$ remain the same. 
Eventually $E_{2,1}$ catches $E_{0,2}$ up, and the energy level
$E_{0,2}=E_{2,1}$ becomes triply degenerate.
By further lowering $\alpha$,
the configuration in Eq. (\ref{a1}) evolves to
\begin{equation}
E_{0,1}<E_{1,1}<E_{0,2}<E_{2,1}<\ldots .
\label{a2}
\end{equation}
Carrying on with the process, as $\alpha$ reaches the  value
(numerically determined) 
%\begin{equation}
$$\alpha_{o}=0.4338\ldots ,$$
%\label{a0}
%\end{equation}
Eq. (\ref{a2}) changes to 
\begin{equation}
E_{0,1}<E_{1,1}=E_{0,2}<E_{2,1}<\ldots,
\label{a3}
\end{equation}
where the energy level $E_{0,2}=E_{1,1}$ is again triply degenerate.
When $\alpha<\alpha_{o}$, Eq. (\ref{a3}) is replaced by
\begin{equation}
E_{0,1}<E_{0,2}<\ldots<E_{1,1}<\ldots.
\label{a4}
\end{equation}
According to Eq. (\ref{einequality}) and the discussion that follows it, 
as $\alpha\rightarrow 0$ 
more and more levels $E_{0,s}$ appear on the left of 
$E_{1,1}$ in Eq. (\ref{a4}), resulting in that the lower part of the 
energy spectrum corresponds to states with vanishing 
angular momentum.

These considerations are relevant to finding the 
behavior of the specific heat, as will be seen in the next section.

\section{Specific heat}

In quantum statistical mechanics \cite{pat72} the internal energy is given by
Eq. (\ref{ienergy}) with the partition function now obtained by
summing over the stationary states discussed in the previous
section,
\begin{equation}
Z=\sum_{n=-\infty}^{\infty}\sum_{s=1}^{\infty}
\exp\{-E_{n,s}/\kappa T\}.
\label{qpfunction}
\end{equation}
As usually happens, a closed expression (in terms of  known analytic functions)
is not available for the
summations in Eq. (\ref{qpfunction}), and  a combination of analytical and
numerical procedures is required to deal with them.
It is convenient to define two characteristic temperatures, namely:
\begin{equation}
\Theta_{\alpha}:= \frac{E_{1,1}-E_{0,1}}{\kappa}
\hspace{2.0cm}
\Theta:= \frac{E_{0,2}-E_{0,1}}{\kappa}
\label{ctemperatures}
\end{equation}
It follows from the discussion in Sec. III that $\Theta_{\alpha}$
increases as $\alpha$ decreases from unity, 
%\begin{equation}
$\Theta_{\alpha=\alpha_{o}}=\Theta$,
%\label{ectemperatures}
%\end{equation}
and as $\alpha\rightarrow 0$,
\begin{equation}
\Theta_{\alpha}\simeq \frac{\hbar^{2}}{2Ma^{2}\kappa\ \alpha^{2}}.
\label{alctemperature}
\end{equation}

As seen in the previous section, the degree of degeneracy of the 
first excited energy level is $2$, $3$ or $1$, for
$\alpha>\alpha_{o}$,  $\alpha=\alpha_{o}$ and $\alpha<\alpha_{o}$,
respectively. Taking this fact to Eq. (\ref{qpfunction}),
Eqs. (\ref{ienergy}) and (\ref{sheat}) yield for the low temperature
behavior ($T\rightarrow 0$) of the specific heat $C(\alpha)$,
\begin{eqnarray}
C(\alpha>\alpha_{o})\simeq 
2\kappa\left(\frac{\Theta_{\alpha}}{T}\right)^{2}e^{-\Theta_{\alpha}/T}
\hspace{0.9cm}
C(\alpha=\alpha_{o})\simeq 
3\kappa\left(\frac{\Theta}{T}\right)^{2}e^{-\Theta/T}
\hspace{0.9cm}
C(\alpha<\alpha_{o})\simeq 
\kappa\left(\frac{\Theta}{T}\right)^{2}e^{-\Theta/T}&&
\label{ltc}
\end{eqnarray}
where the integers multiplying $\kappa$ are the degree of degeneracy
just mentioned.
Eq. (\ref{ltc}) shows that $C$ varies discontinuously 
with $\alpha$ when $T$ is much smaller than the 
characteristic temperatures in Eq. (\ref{ctemperatures}). 
As $T$ increases, higher energy levels have to be
considered in Eq. (\ref{qpfunction}), and Eq. (\ref{ltc}) no longer holds.

\begin{figure}[thp]
\leavevmode
\centering
\includegraphics[scale=0.8]{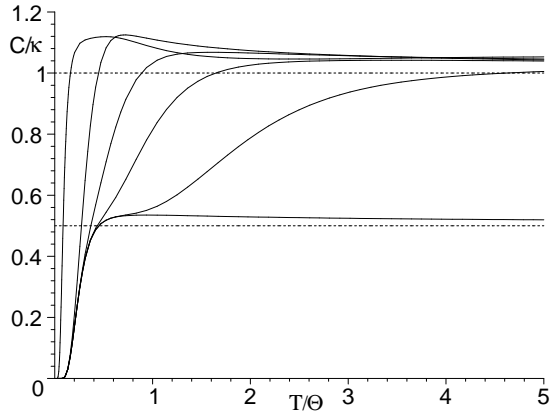}
\caption{Numerical plots of $C/\kappa$ versus $T/\Theta$.
The plots from the left correspond to 
$\alpha$=1,0.3,0.2,0.15,0.1 and 
$\alpha\rightarrow 0$, respectively. 
The plots for $\alpha$=1 to $\alpha$=0.15 display 
maxima corresponding to specific heat  
of about 1.119 $\kappa$, 1.125 $\kappa$, 1.068 $\kappa$ and 1.042 $\kappa$ at
temperatures of about 
0.52 $\Theta$, 0.72 $\Theta$, 1.59 $\Theta$ and 3.49 $\Theta$, respectively.
At $T\simeq 0.94\ \Theta$,
the plot for $\alpha\rightarrow 0$ displays a maximum 
corresponding to $C\simeq 0.535\ \kappa$.
\label{econe}
}
\end{figure}

The way that $C$ departs from Eq. (\ref{ltc}) is shown in Fig. \ref{econe}.
In particular, for small values of $\alpha$
the ways that $C$ departs from $C(\alpha<\alpha_{o})$ in
Eq. (\ref{ltc}) are approximately the same. 
This can be understood by recalling that when $\alpha$ is small the lower 
part of the spectrum is essentially formed by energy levels $E_{0,s}$ 
which do not depend on $\alpha$.
Eq. (\ref{alctemperature}) shows that for small $\alpha$
the azimuthal characteristic temperature 
$\Theta_{\alpha}$ is high, and  consequently
the various plots of $C$ split from each other only
at higher temperatures, when the populations of  
energy levels with nonvanishing angular momentum become appreciable.

At the regime $T\ll\Theta_{\alpha}$ with $\alpha<\alpha_{o}$
[$\alpha\geq \alpha_{o}$ necessarily leads to the
corresponding low temperature behavior in Eq. (\ref{ltc})],
only states with $n=0$  effectively counts
in  Eq. (\ref{qpfunction}), resulting
\begin{equation}
Z\simeq\sum_{s=1}^{\infty}
\exp\{-E_{0,s}/\kappa T\}.
\label{qpfunctionn0}
\end{equation}
By setting $\nu=0$ in Eq. (\ref{avalue}), Eq. (\ref{energy}) yields
\begin{equation}
E_{0,s}=\frac{h^{2}}{8Ma^{2}}\left(s-\frac{1}{4}\right)^{2} + \ldots,
\label{energyn0}
\end{equation}
which up to the term 1/4 and the corrections
is the energy of a free particle in 
a one-dimensional box of length $a$ 
(and this identification improves as $s$ increases).
As $T\rightarrow 0$, Eq. (\ref{qpfunctionn0}) leads to 
$C(\alpha<\alpha_{o})$ in Eq. (\ref{ltc}). When $\alpha$ is sufficiently small
such that 
\begin{equation}
\Theta\ll T\ll\Theta_{\alpha}
\label{ltregime}
\end{equation}
applies, the summation in Eq. (\ref{qpfunctionn0}) can be replaced, 
in first approximation,
by an integration over $s$ from 1 to $\infty$, with $E_{0,s}$
given by Eq. (\ref{energyn0}) [one can safely truncate Eq. (\ref{energyn0})
at the first term to obtain the main contribution]. 
Using for example Refs. \cite{abr65} or \cite{pru86},
evaluation of the integration yields $Z\simeq Z_{(1)}$,
where 
\begin{equation}
Z_{(1)}=\frac{a}{h}\sqrt{2\pi M\kappa T}
\label{cpfunction1}
\end{equation}
is the classical partition function 
of a particle in a one-dimensional box of length $a$.
Eqs. (\ref{ienergy}) and (\ref{sheat}) then give 
$U\simeq\kappa T/2$ and 
\begin{equation}
C\simeq\frac{\kappa}{2},
\label{firstc}
\end{equation}
showing that the conical singularity indeed
suppresses one degree of freedom [cf. Eq. (\ref{ienergy2})].

With $\alpha$ smaller than about $0.1$, 
Fig. \ref{econe} shows that $C$ gets a maximum value
just above $\kappa/2$, at $T$ close to $\Theta$
(see Fig. \ref{econe} for more accurate values).
If $\Theta\ll\Theta_{\alpha}$ also
holds, the regime $T\ll\Theta_{\alpha}$ resembles 
the thermal behavior of the specific heat of a particle in a one-dimensional
box of length $a$ \cite{ron62}. 

When the temperature is much larger than both the characteristic temperatures,
one can proceed as above and replace the summations in
Eq. (\ref{qpfunction}) by integrations. Noting 
Eqs. (\ref{states}) to (\ref{energy}), and 
performing the integration
over $s$ \cite{abr65,pru86} leads to an integration over $n$ 
of a complementary error function, namely
$$\int_{0}^{\infty} {\rm erfc}(x)\ dx=\frac{1}{\sqrt{\pi}},$$
which has been evaluated by regularizing
related  formulae in Refs. \cite{abr65,gra80}. 
The leading contribution in
$Z$ is proportional to the classical partition function in
Eq. (\ref{cpfunction2}), with the proportionality constant depending on where
Eq. (\ref{avalue}) is truncated: keeping only the first term, and then
also the second term, results 
\begin{equation}
Z\simeq 0.81\ Z_{(2)}\hspace{2.5cm}Z\simeq 0.94\ Z_{(2)}
\label{pfunctions}
\end{equation}
respectively
(in fact, to obtain the second expression 
one has to evaluate an additional integration involving the complementary
error function \cite{pru86}).
Using either results in Eq. (\ref{pfunctions}),
Eqs. (\ref{ienergy}) and (\ref{sheat}) yield as leading contribution 
\begin{equation}
C\simeq\kappa
\label{secondc}
\end{equation}
which is the classical value, as expected.
[The discrepancies in Eq. (\ref{pfunctions}) 
and its failure in reproducing $Z_{(2)}$ exactly
can be understood by noticing that for
large $n$ in Eq. (\ref{qpfunction}), terms beyond the first term in 
Eq. (\ref{avalue}) become relevant.]

A maximum also appears, now above $\kappa$,
at a temperature comparable to $\Theta_{\alpha}$,
as shown in Fig. \ref{econe}.
In particular the plot for $\alpha=1$ (particle in a flat disk)
resembles that of a particle  in a square box of the same area \cite{ron62}.

%\begin{widetext}
%\end{widetext}

\section{Conclusion}
Motivated by the fact that statistical mechanics of systems in 
atomic size  boxes is very sensitive to the topology and geometry
of the boxes,
this work treated the 
specific heat $C$ of a free nonrelativistic particle in a conical box.
The energy spectrum
was determined and its features were obtained from properties
of the zeros of the Bessel functions. In particular, degeneracy
of the energy levels was related to long established theorems regarding common zeros
of the Bessel functions.

This toy model exposed with clarity the effects of a
conical singularity on $C$ at various regimes of temperature.
At low temperatures $C$ varies discontinuously with the deficit angle
[cf. Eq. (\ref{ltc})]. 
At low and intermediate temperatures, 
if the cone is sufficiently sharp,
the azimuthal degree of freedom is suppressed [cf. Eq. (\ref{firstc})]. 
At high temperatures
the radial and the azimuthal degrees of freedom are fully excited,
leading to the value of $C$ prescribed by the equipartition theorem  
[cf. Eq. (\ref{secondc})]. The numerical plots in Fig. \ref{econe}
illustrate these facts.

The effect of the sharpness of the cone on $C$ can be better appreciated
by contrasting the two limits $\alpha=1$ and $\alpha\rightarrow 0$.
When $\alpha=1$, the energy spectrum is that of a free particle in a flat
disk, where energy levels with vanishing and nonvanishing angular momentum 
are mixed up along the spectrum. 
As temperature $T$ rises,
the radial and the azimuthal degrees of freedom get equally excited, leading
to the high temperature value $C\simeq\kappa$, 
which is consistent with the equipartition theorem. 
This picture changes radically as one sets $\alpha\rightarrow 0$.
When the limit $\alpha\rightarrow 0$ is literally taken, 
the energy of the first excited state with nonvanishing angular
momentum gets infinitely  high in the spectrum. Then, as 
$T$ increases, only the radial degree of freedom gets progressively excited,
resulting in $C\simeq\kappa/2$ at higher temperatures ---
the  $\alpha\rightarrow 0$ limit suppresses the azimuthal degree of freedom.

This work also aimed to tune mathematical tools
in quantum statistical mechanics on the cone
such that other thermodynamical quantities, as well as more 
elaborate models, can be addressed. 
Some remarks are in order.

The regular solutions in Eq. (\ref{states}) are not the only ones consistent
with square integrability of the wave function and conservation of probability
\cite{kay91,hel01}. A mild divergence at the origin 
(where the conical singularity sits) 
does not spoil these physical requirements \cite{mor98}. 
It seems that the choice
for regular solutions relies in that they arise naturally when the conical
singularity is ``smoothed" by regularization \cite{kay91}.

The discontinuity of  the specific heat in Eq. (\ref{ltc})
is a feature of any thermal system whose degree of degeneracy $g$
of the first excited energy level $\epsilon _{1}$ varies with
some parameter. If $\epsilon _{o}$ is the zero point energy,
as $T\rightarrow 0$, a quick calculation shows that  
$$C\simeq 
g\kappa\left(\frac{\theta}{T}\right)^{2}e^{-\theta/T}$$
where $\theta:=(\epsilon_{1}-\epsilon_{o})/\kappa$.
This kind of discontinuity at low temperatures also appears,  
for example,  in the specific heat of the anisotropic rigid rotator \cite{car84}.

Recalling that even the main contributions in Eq. (\ref{pfunctions})
were affected by where one truncates Eq. (\ref{avalue}),
the determination of corrections in 
Eqs. (\ref{firstc}) and (\ref{secondc}) turns out to
be a rather tricky task. Although the leading contributions in
Eqs. (\ref{firstc}) and (\ref{secondc}) do not depend
on where Eq. (\ref{avalue}) is truncated
(either on the lower limit of the related integrations), 
this may not be the case when dealing with their corrections.
Such a problem is somewhat similar to that where two distinct 
approximations for the energy spectrum give  the  same
leading  contribution in the expression for the high temperature 
behavior of $C$; but different corrections
(e.g., the anharmonic oscillator in Ref. \cite{mar84}). 
Therefore, in order to check correctness at every step
in using Eq. (\ref{avalue}) to finding subleading contributions, 
a thorough numerical analysis 
must be implemented 
(it should be noted that this is the approach in Ref. \cite{mar84}).
 
Since the disk and the cone have the same topology \cite{mor98},
the suppression of one degree of freedom in Eq. (\ref{firstc})
is a geometrical rather than a topological effect --- it is due to
a curvature singularity.

Although the background addressed in this work is a cone,
for which $\alpha$ is not greater than unity [cf. Eq. (\ref{cparameter})],
it should be pointed out that 
in the context of topological defects in solids \cite{kro81}
there is room for $\alpha>1$ in the effective geometry
corresponding to disclinations.
In this case, as $\alpha$ increases from unity,  $E_{1,1}$ approaches $E_{0,1}$
and $\Theta_{\alpha}$ in Eq. (\ref{ctemperatures}) decreases, suggesting that
the classical limit in Eq. (\ref{secondc}) would be 
reached at lower temperatures (this point needs further investigation though).

\begin{acknowledgments}
The authors are grateful to Renato Klippert and Ricardo Medina for their
assistance on computational matters.
This work was partially supported by the 
research agencies CNPq and FAPEMIG.

\end{acknowledgments}

%\end{multicols}

\end{document}